\documentclass[aps,onecolumn,prd,showpacs,nofootinbib]{revtex4}
\usepackage{amsmath}
\usepackage{graphicx}
\usepackage{dcolumn}
\usepackage{bm}
\usepackage{amssymb}
\usepackage{latexsym}

\def\be{\begin{equation}}
\def\ee{\end{equation}}
\def\ba{\begin{eqnarray}}
\def\ea{\end{eqnarray}}

\begin{document}

\title{\Large{G-Curvaton} \vspace{6mm}}

\author{Hua Wang$^1$}
\email{wanghua09@mails.gucas.ac.cn}

\author{Taotao Qiu$^{2,3,4}$}
\email{xsjqiu@gmail.com}

\author{Yun-Song Piao$^1$}
\email{yspiao@gucas.ac.cn}

\vspace{16mm}

\affiliation{$1$ College of Physical Sciences, Graduate University
of Chinese Academy of Sciences, Beijing 100049, China \vspace{1mm}}
\affiliation{$2$. Department of Physics and Center for Theoretical
Sciences, National Taiwan University, Taipei 10617, Taiwan}
\affiliation{$3$. Leung Center for Cosmology and Particle
Astrophysics National Taiwan University, Taipei 106, Taiwan}
\affiliation{$4$. TPCSF, Institute of High Energy Physics, CAS, P.O.
Box 918-4, Beijing 100049, P.R. China}

\begin{abstract}
\vspace{3mm}
\begin{center} \textbf{Abstract}\vspace{3mm}
\end{center}

In this paper, we study a curvaton model where the curvaton is acted
by Galileon field. We calculate the power spectrum of fluctuation of
G-curvaton during inflation and discuss how it converts to the
curvature perturbation after the end of inflation. We estimate the
bispectrum of curvature perturbation induced, and show the
dependence of non-Gaussianity on the parameters of model. It is
found that our model can have sizable local and equilateral
non-Gaussianities to up to ${\cal O}(10^2)$, which is illustrated by
an explicit example.

\end{abstract}


\maketitle

\section{Introduction}
Inflation has now been considered as one of the most successful
theory to describe our universe at early epoches
\cite{Guth:1980zm,Albrecht:1982wi,Linde:1983gd}. Making our universe
expand fast enough, it can naturally solve many notorious problems
brought out by hot Big Bang, such as flatness problem, horizon
problem, monopole problem and so on. Moreover, during inflation the
quantum fluctuations generated at the initial stage can be stretched
out of the horizon to form classical perturbations, which can
provide seeds for the formation of the structure of our universe. An
inflation model also succeeds in producing nearly scale-invariant
power spectrum of scalar perturbation and tiny gravitational waves,
which fits very well with today's observational data
\cite{Larson:2010gs}. The non-Gaussian corrections of perturbations
during inflation can also be large or not, according to various
inflation models, which is waiting for constraints from new and more
accurate data in the near future \cite{:2006uk}.

Usually, the perturbation of inflation is generated by the inflaton
field itself, which is the simplest way to have curvature
perturbation. However, it is not the only choice. Perturbations
generated in such a way depends on the potential of the inflaton
field, and thus puts very severe constraint on inflation models. In
order to relax such a constraint, as Lyth and Wands have pointed
out, perturbation can also be generated from another field that has
nothing to do with the inflaton field, namely, the {\it curvaton}
\cite{Lyth:2001nq}, see also relevant works on curvaton mechanism in
Refs. \cite{Enqvist:2001zp, Moroi:2001ct} and earlier
\cite{Mollerach, Linde:1996gt}. Curvaton field is usually assumed to
be a scalar field with light mass and decoupled from all the other
kinds of perturbations, thus the perturbation produced by curvaton
can be independent on the nature of inflation. Moreover, since the
curvaton is subdominant during inflation, it can only produce
isocurvature perturbation. This isocurvature perturbation has to be
converted into curvature perturbation at the end of inflation, so it
depends on what happened after the inflation terminated. Usually,
there are two cases in which this conversion can be available:
First, when the inflaton decays into radiation after inflation, the
curvaton field becomes dominate; Second, the curvaton field decays
as well before its domination, and reaches equilibrium with
radiation that decays from inflaton. According to different case,
the amount of the curvature perturbation converted from curvaton
perturbation may be different. Curvaton scenarios have been widely
studied in, for instance, \cite{Sasaki:2006kq}.

In the original curvaton paper, it was suggested that curvaton is
made of a canonical scalar field. However, other field models can be
considered to act as a curvaton. People have considered curvaton of
Pseudo-Nambu-Goldstone-Boson \cite{Dimopoulos:2003az}, DBI-type
\cite{Li:2008fma} or its curvaton brane implement
\cite{Zhang:2009gw}, multi-field \cite{Huang:2008rj}, lagrangian
multiplier field \cite{Feng:2010tf}, Horava-Lifshitz scenario
\cite{Mukohyama:2009gg} and so on. Very recently, a new kind of
models has been proposed and studied extensively, which is called
``Galileon" models \cite{Nicolis:2008in}. The original version of
these models is a generalization of an effective field description
of the DGP model \cite{Dvali:2000hr}. These models includes high
derivative operator of the scalar field, however, due to some
``delicated design", its equation of motion remains second order,
\footnote{This idea is actually pioneered by Horndeski thirty years
ago, in \cite{Horndeski:1974}.} thus it can violate the NEC without
incorporating any instability modes. Due to such an interesting
property, such kind of model generically can admit superluminal
propagation \cite{Nicolis:2009qm} and perhaps even closed timelike
curves (CTCs) \cite{Evslin:2011vh} (see also \cite{Babichev:2007dw}
for CTCs without Galileon). Moreover, these models can be applied
onto various evolution period of our universe, such as dark energy
\cite{Deffayet:2010qz, Silva:2009km}, inflation
\cite{Kobayashi:2010cm, Creminelli:2010ba, Burrage:2010cu},
reheating \cite{LevasseurPerreault:2011mw}, bouncing
\cite{Qiu:2011cy}, the slow expansion scenario \cite{Piao:2003ty} of
primordial universe \cite{Liu:2011}, \cite{Piao:2011mq}, and so on.
As an extension, these models can also be generalized to DBI version
\cite{deRham:2010eu}, the K-Mouflage scenario
\cite{Babichev:2009ee}, the supersymmetric Galileon
\cite{Khoury:2011da}, the Kinetic Gravity Braiding models
\cite{Deffayet:2010qz}, the generic Galileon-like action
\cite{Horndeski:1974, Deffayet:2011gz, Gao:2011qe} and others. See
also e.g. Refs. \cite{Deffayet:2009wt, Chow:2009fm, Mizuno:2010ag,
Nesseris:2010pc, DeFelice:2011zh, Padilla:2010ir,
Hinterbichler:2010xn, Creminelli:2010qf} for various study of their
phenomenologies.

In the present work, we study the scenario where a Galileon field
behaves as a curvaton, which we dubbed as ``{\it G-Curvaton}"
scenario. Due to the higher derivative term, Galileon is expected to
have some features on generating perturbations, such as getting
large tensor-scalar ratio in ``G-inflation" scenario. Our paper is
organized as follows: in Sec. II we review the original curvaton
scenario, by taking curvaton to be the simplest one, i.e., the
canonical scalar field. In Sec. III we study our ``G-Cuvaton" model.
We first investigate its perturbation, obtaining a scale-invariant
power spectrum. Then we discuss how it converts to the curvature
perturbation at the end of inflation, considering both cases where
curvaton decays before and after it dominates the universe. We also
study the non-Gaussianities generated from our model, both local
type and equilateral type. Finally we present an explicit example to
show how the observable quantities could be effected. Sec. IV comes
our conclusion and discussions.

\section{Review of the Simplest Curvaton Model}
In this section, we would like to review how the mechanism works for
the simplest curvaton model, which is made of a canonical scalar
field. The Lagrangian for the curvaton field is:
\begin{eqnarray}\label{lagrangian} {\cal
L}_\sigma=-\frac{1}{2}\partial_\mu\sigma\partial^\mu\sigma-V(\sigma)~.
\label{Lsimple}\end{eqnarray} Note that here we are using the metric
with notation $ds^2 = - dt^2+a^2(t)\delta_{ij}dx^idx^j$. As for
curvaton, the effective mass of $\sigma$ needs to be very light,
which put the constraint $|V_{,\sigma\sigma}|\ll H^2$ to the
potential $V(\sigma)$, where
$V_{,\sigma\sigma}\equiv\partial^2V/\partial\sigma^2$ and $H$
denotes the Hubble parameter of the universe. Moreover, during
inflation $H$ is almost a constant. Thus in general, one may define
the parameters
\begin{eqnarray} \epsilon\equiv -\frac{\dot H}{H^2}~,~ \xi={V_{,\sigma\sigma}\over
3H^2}, \end{eqnarray} which are required to be far smaller than 1
and slowly varying. Moreover, since curvaton is subdominant part of
the universe during inflation, the energy scale of the potential
should also be lower than that of inflation, namely $V(\sigma)\ll
3M_{pl}^2H^2$.

Suppose the field generates fluctuation during inflation, namely
$\sigma(x)\rightarrow\sigma_0(t)+\delta\sigma(t,{\bf x})$, then from
Eq. (\ref{lagrangian}) we can easily get the equation of motion for
the field fluctuation $\delta\sigma$ as:
\begin{eqnarray}\ddot{\delta\sigma}_{\bf k}+3H\dot{\delta\sigma}_{\bf
k}+((k/a)^2+V_{,\sigma\sigma})\delta\sigma_{\bf k}=0~,
\end{eqnarray} where $\delta\sigma_{\bf k}$ is the Fourier
presentation of $\delta\sigma$ with momentum mode ${\bf k}$, and $a$
is the scale factor. Note that here we have already neglected the
coupling of $\delta\sigma_{\bf k}$ to the metric perturbation, as
has been done in \cite{Lyth:2001nq}.
Thus the power spectrum of the field fluctuation $\delta\sigma_{\bf
k}$ at the horizon crossing is given by
\begin{eqnarray}\label{spectrum} {\cal
P}_\sigma\equiv\frac{k^3}{2\pi^2}|\delta\sigma_{\bf
k}|^2=\bigg(\frac{H_\ast}{2\pi}\bigg)^2~,\end{eqnarray} where the
star denotes the time of horizon exit, $k=a_\ast H_\ast$. The
spectral index of the spectrum can also be given by:
\begin{eqnarray} n_\sigma-1\equiv\frac{d\ln {\cal
P}_\sigma}{d\ln k}=-2\epsilon+2\xi\ll
1~,\label{nsimple}\end{eqnarray} which shows the nearly
scale-invariance of the spectrum ${\cal P}_\sigma$.

However, since the curvaton is the subdominant part of the universe
during inflation, the perturbation generated by it can only be of
isocurvature type. This type of perturbation can only be converted
into curvature perturbation when curvaton dominates, or become
equilibrium with other part of the universe. Either these cases will
not happen during inflation, however after inflation, when inflaton
field decays into radiation whose energy density may decrease more
rapidly than curvaton field, both the two cases will happen,
depending on when the curvaton field will decay. In the case when
the curvaton decays late, it will exceed over the radiation decayed
from inflaton and dominate the universe, however in the case when
the curvaton decays early, it may become equilibrium with the
radiation. The curvature perturbations for a given matter with
energy density $\rho$ in spatial-flat slicing is given by
\cite{Wands:2000dp}:
\begin{eqnarray}
\zeta=-H\frac{\delta\rho}{\dot\rho}~,
\end{eqnarray} and the separately conserved curvature perturbations
for radiation and curvaton field therefore read: \begin{eqnarray}
\label{radiationcurv}\zeta_r&=&-H\frac{\delta\rho_r}{\dot\rho_r}=\frac{1}{4}\frac{\delta\rho_r}{\rho_r}~,\\
\label{curvatoncurv}\zeta_\sigma&=&-H\frac{\delta\rho_\sigma}{\dot\rho_\sigma}=~\frac{\delta\rho_\sigma}{3(\rho_\sigma+p_\sigma)}~,
\end{eqnarray} respectively, where $p_\sigma$ is the pressure of the
curvaton field. Using these results, and assuming that the
isocurvature perturbation convert to curvature perturbation
instantly, then the curvature perturbation generated by such
conversion reads:
\begin{eqnarray}\label{curvature}
\zeta=\frac{4\rho_r\zeta_r+3(\rho_\sigma+p_\sigma)\zeta_\sigma}{4\rho_r+3(\rho_\sigma+p_\sigma)}\simeq\frac{3(\rho_\sigma+p_\sigma)\zeta_\sigma}{4\rho_r+3(\rho_\sigma+p_\sigma)}~,
\end{eqnarray} where in the last step we neglected the curvature perturbation of
radiation, $\zeta_r$. Define the energy density ratio of $\sigma$
over radiation, $r\equiv\rho_\sigma/\rho_r$, then in the first case
where the curvaton dominates, we have $r\gg 1$, Eq.
(\ref{curvature}) becomes: \begin{eqnarray}\label{curvature1}
\zeta\simeq\zeta_\sigma~,
\end{eqnarray} while in the second case where $\sigma$ become
equilibrium with radiation, we have $\rho_\sigma\ll\rho_r$, Eq.
(\ref{curvature}) becomes:
\begin{eqnarray}\label{curvature2} \zeta\simeq\frac{3}{4}r(1+w_\sigma)\zeta_\sigma~,
\end{eqnarray} where $w_\sigma\equiv p_\sigma/\rho_\sigma$ is the
equation of state of the curvaton field.

Furthermore, we investigate the non-Gaussianities of the
perturbation generated by the curvaton field. The local type
non-Gaussianities of curvature perturbation are given by:
\begin{eqnarray}\label{zeta}
\zeta=\zeta_g+\frac{3}{5}f^{local}_{NL}\zeta_g^2~,\end{eqnarray}
where the subscript ``g" denotes the Gaussian part of $\zeta$ while
$f_{NL}$ is the so-called nonlinear estimator. For local type,
$f^{local}_{NL}$ can be estimated by using the so-called $\delta N$
formalism, where $\delta N$ is the variation of the number of efolds
$N$ of inflation \cite{Starobinsky:1986fxa}:
\begin{eqnarray}\label{deltan} \zeta=\delta
N=N_{,\sigma}\delta\sigma+\frac{1}{2}N_{,\sigma\sigma}\delta\sigma^2+...~,
\end{eqnarray} where
$N_{,\underbrace{\sigma...\sigma}_n}\equiv\partial^n
N/\partial\sigma^n$ and the same notations hereafter. Comparing Eqs.
(\ref{zeta}) and (\ref{deltan}) one can easily find that
$f^{local}_{NL}$ can be presented using $N_{,\sigma}$ and
$N_{,\sigma\sigma}$, namely,
\begin{eqnarray}
f^{local}_{NL}\Big|_\zeta=\frac{5}{6}\frac{N_{,\sigma\sigma}}{N_{,\sigma}^2}~.
\end{eqnarray}

For nonlocal type, however, things will become a little bit more
complicated, since non-Gaussianities also exists in the field
fluctuation $\delta\sigma$ itself, and $\delta N$ formalism will not
be valid any longer. In this case, we could express the 3-point
correlation function of $\zeta$ as:
\begin{eqnarray}\label{nonlocal1}
\langle|\zeta(k_1)\zeta(k_2)\zeta(k_3)|\rangle=(2\pi)^{3}\delta^{3}(\sum_{i}k_{i})\mathcal{B}(k_{1},k_{2},k_{3})
\end{eqnarray} where $\mathcal{B}(k_{1},k_{2},k_{3})$ is the shape
of the non-Gaussianties and
\begin{eqnarray}\label{nonlocal2}
\langle|\zeta(k_1)\zeta(k_2)\zeta(k_3)|\rangle=-i\mathcal{T}\int_{t_{0}}^{t}dt^{\prime}\langle|[\zeta(t,k_{1})\zeta(t,k_{2})\zeta(t,k_{3}),{\cal
H}_{int}^p(t^{\prime})]|\rangle~, \end{eqnarray} with ${\cal
H}_{int}^p$ being interaction Hamiltonian of curvaton in momentum
space. The nonlinear estimator is defined as:
\begin{eqnarray}\label{nonlocal3} f^{nonlocal}_{NL}\Big|_\zeta\equiv
\frac{10}{3}\frac{\prod_{i=1}^3k_i^3}{(2\pi)^4\sum_{i=1}^3k_i^3}\frac{\mathcal{B}(k_{1},k_{2},k_{3})}{[{\cal
P}_\zeta(k_1){\cal P}_\zeta(k_2)+2perm.]}~. \end{eqnarray}

\section{The G-curvaton model}
In this section, we study our model of which the curvaton field is
of Galileon type. For simplicity but without losing generality, we
consider the Lagrangian of curvaton has the general third order
Galileon field, which is given by \cite{Nicolis:2008in,
Deffayet:2010qz}:
\begin{eqnarray}\label{actiongc}
{\cal S}_{GC}=\int
d^4x\sqrt{-g}[K(\sigma,X)-G(\sigma,X)\Box\sigma]~,
\end{eqnarray}
where $K$ and $G$ are generic functions of $\sigma$ and
$X\equiv-\partial_\mu\sigma\partial^\mu\sigma/2$ is the kinetic
term of the field $\sigma$. The Lagrangian used is inspired by the
original Galileon construction but does not respect the Galileon
symmetry, which was called `Generalised Galileons' by Deffayet et
al.\cite{Deffayet:2010qz}. Note that more generalized Galileon
model containing higher order operators of $\Box\sigma$ or the
couplings of $\sigma$ to the gravitational part was constructed in
Ref. \cite{Deffayet:2011gz}. From action (\ref{actiongc}), the
energy-momentum tensor $T_{\mu\nu}$ has the form of:
\begin{eqnarray}
T_{\mu\nu}=K_{,X}\partial_\mu\sigma\partial_\nu\sigma+K
g_{\mu\nu}-\partial_{\mu}G \partial_{\nu}\sigma-\partial_{\mu}G
\partial_{\nu}\sigma + g_{\mu\nu}\partial_{\lambda
}G\partial^{\lambda }\sigma
-G_{,X}\Box\sigma\partial_\mu\sigma\partial_\nu\sigma~.
\end{eqnarray}
Taking the homogeneous and isotropic background where $T_{\mu\nu}$
has the form of $diag\{\rho_\sigma, a^2(t)p_\sigma, a^2(t)p_\sigma,
a^2(t)p_\sigma\}$, we can furtherly obtain the energy density and
pressure of $\sigma$ as:
\begin{eqnarray}\label{rho}
\rho_\sigma&=&2K_{,X}X-K+3HG_{,X}\dot\sigma^3-2G_{,\sigma} X~,\\
\label{p} p_\sigma&=&K-2\left(G_{,\sigma}+G_{,X}\ddot\sigma
\right)X~.
\end{eqnarray}
Moreover, by varying action (\ref{actiongc}) with respect to
$\sigma$, we obtain the equation of motion for $\sigma$:
\begin{eqnarray}
&&K_{,X}\Box\sigma-K_{,XX}(\partial_{\mu}\partial_\nu\sigma)(\partial^\mu\sigma\partial^\nu\sigma)
-2K_{,X\sigma}X+K_{,\sigma} -2\left( G_{,\sigma}-
G_{,X\sigma}X\right)\Box\sigma \nonumber\\&& +G_{,X}\left[
(\partial_\mu\partial_\nu\sigma)(
\partial^\mu\partial^\nu\sigma)-(\Box\sigma)^2+R_{\mu\nu}\partial^\mu\sigma\partial^\nu\sigma
\right]
+2G_{,X\sigma}(\partial_{\mu}\partial_\nu\sigma)(\partial^\mu\sigma\partial^\nu\sigma)
\nonumber\\&& +2G_{,\sigma\sigma}X-G_{,XX}\left(
\partial^\mu\partial^\lambda\sigma -g^{\mu\lambda}\Box\sigma
\right)\left(\partial_\mu\partial^\nu\sigma\right)\partial_\nu\sigma\partial_\lambda\sigma
=0~.
\end{eqnarray}
which can be simplified in FRW universe where the background are
homogeneous and isotropic:
\begin{eqnarray}\label{bgeom}
&& K_{,X} \left( \ddot{\sigma}+3H\dot{\sigma} \right) + 2 K_{,XX} X
\ddot{\sigma} +
   2K_{,X\sigma} X - K_{,\sigma}
-2\left( G_{,\sigma}-G_{,X\sigma}X \right)\left(
\ddot{\sigma}+3H\dot{\sigma} \right) \nonumber\\&& + 6G_{,X} \left[
\left( HX \right)\dot{}+3H^2 X \right] - 4G_{,X\sigma} X
\ddot{\sigma} - 2G_{,\sigma\sigma}X+6HG_{,XX}X \dot{X}=0~.
\end{eqnarray}

\subsection{Scale-Invariant Power Spectrum for Field Fluctuation}

To study the perturbations of the curvaton field, we can split the
scalar field $\sigma$ into
$\sigma(x)\rightarrow\sigma_0(t)+\delta\sigma(t, {\bf x})$, where
$\sigma_0$ represents the spatially homogeneous background field,
and the $\delta\sigma$ stands for the linear fluctuation which
corresponds to the isocurvature perturbation during inflation.
Taking the spatial-flat gauge and using the assumption that there is
no coupling between $\delta\sigma$ and other perturbations, one can
have the equation of motion for field fluctuation in momentum space
$\delta\sigma_{\bf k}$ as:
\begin{eqnarray}\label{eom}
\ddot{\delta\sigma}_{\bf k}+ \bigg(3+\frac{\dot{{\cal D}}}{H{\cal
D}} \bigg)H \dot{\delta\sigma}_{\bf k} + \frac{c_s^2k^2}{a^2}
\delta\sigma_{\bf k} +{\cal M}_{eff}^2\delta\sigma_{\bf k}=0~,
\end{eqnarray}
where we have defined $c_s^2\equiv{\cal C}/{\cal D}$, with
\begin{eqnarray}
\label{c}{\cal
C}&=&K_{,X}+2G_{,X}\left(\ddot\sigma_0+2H\dot\sigma_0\right)
+2G_{,XX}X\ddot\sigma_0-2\left(G_{,\sigma}-G_{,X\sigma}X\right)~,\\
\label{d}{\cal D}&=&K_{,X}+2K_{,XX}X+6HG_{,X}\dot\sigma_0
-2\left(G_{,\sigma}+G_{,X\sigma}X\right)+6HG_{,XX}X\dot\sigma_0~.
\end{eqnarray} Note that in order to avoid ghost or gradiant
instabilities in our model, one must require $\mathcal{C}\geq 0$,
$\mathcal{D}>0$, which leads to the non-negativity of the sound
speed squared: $c_s^2\geq 0$. The effective mass squared is:
\begin{eqnarray}\label{mass}
{\cal
M}_{eff}^{2}&=&\frac{1}{a^{3}}\frac{d}{dt}\big[a^{3}(K_{,X\sigma}\dot{\sigma}_{0}+3HG_{,X\sigma}\dot{\sigma}_{0}^{2}-G_{,\sigma\sigma}\dot{\sigma}_{0})\big]
-K_{,\sigma\sigma}+G_{,\sigma\sigma}\Box\sigma_{0}~.
\end{eqnarray}
For later convenience, we also introduce the following ``slow
variation" parameters:
\begin{eqnarray}\label{slowroll1}
\epsilon_G=-\frac{\dot H}{H^2}~,~ s_G=\frac{\dot{c_s}}{Hc_s}~,~
\delta_G=\frac{\dot{{\cal D}}}{H{\cal D}}~,~\xi_G={{\cal
M}_{eff}^{2}\over 3H^2},
\end{eqnarray} which are assumed to be small but not neglected.
When reduced to the simplest curvaton model given by
(\ref{Lsimple}), $\epsilon_G=\epsilon$, $\xi_G=\xi$ and
$s_G=\delta_G=0$.

For solving Eq. (\ref{eom}) and computing the power spectrum, it is
convenient to turn to the conformal coordinate where conformal time
$\tau$ is defined as $d\tau\equiv dt/a$. Using a new variable
$u_{\bf k}\equiv z\delta\sigma_{\bf k}$ where $z\equiv
a\sqrt{\mathcal{D}}$, the equation of motion can be written in the
Fourier space as
\begin{eqnarray}\label{eom2}
u_k''+\left(c_s^2k^2-\frac{z''}{z}\right)u_k=0~,
\end{eqnarray}
and the prime denotes differentiation with respect to $\tau$. Under
the slow roll approximation where all the parameters in
(\ref{slowroll1}) are small, we find:
\begin{eqnarray}
\frac{z''}{z}\simeq\frac{2}{\tau^2}.
\end{eqnarray}
Thus we finally can obtain the power spectrum of $\delta\sigma_{\bf
k}$ at horizon crossing as (using definition (\ref{spectrum})):
\begin{eqnarray}\label{spectrum2} {\cal
P}_{\delta\sigma_{\bf k}}=\frac{H^2_\ast}{4\pi^2 c_s^3\mathcal{D}}
\end{eqnarray} with the spectral index given by
\begin{eqnarray}\label{index}
n_{\sigma}-1=\frac{d\ln {\cal P}_\sigma}{d\ln k}=
-2\epsilon_G-3s_G-\delta_G+2\xi_G \ll 1.
\end{eqnarray}
From Eq.(\ref{index}) we can see that the isocurvature perturbation
generated by our curvaton field can give rise to a very flat power
spectrum which is nearly scale-invariant. Moreover, comparing to
usual curvaton case (\ref{spectrum}), the amplitude of the power
spectrum is suppressed by Galileon-like nonlinear terms such as
$\cal D$, however, when $G(\sigma,X)=0$ and
$K(\sigma,X)=X-V(\sigma)$, the result exactly reduces to that of
usual curvaton case. When $G(\sigma,X)=0$ and $K(\sigma,X)$ is
DBI-type, the result is that of DBI-curvaton\cite{Li:2008fma}. The
result given here is also consistent with the case where Galileon
field act as an inflaton field and generate curvature perturbations
\cite{Kobayashi:2010cm}\footnote{Strictly speaking, as authors of
Ref. \cite{Kobayashi:2010cm} pointed out themselves, what they
calculated is not usual comoving curvature perturbation but another
variable which coincide with comoving curvature perturbation only in
large scales.}, and the similar property has also been found in
other featured inflation models, such as DBI inflation
\cite{Alishahiha:2004eh}.

\subsection{Generating the Curvature Perturbation from Curvaton}

From last paragraph we learned that our G-Curvaton model is able to
give rise to perturbations with nearly scale-invariant power
spectrum, however these perturbations are of isocurvature ones. As
has been mentioned before, curvature perturbations can be obtained
after inflation, so now we consider the epoch when inflation has
ceased and the inflaton has already decayed to radiation, during
which the universe is filled with the curvaton field $\rho_\sigma$
and the radiation $\rho_r$. From this moment on, the isocurvature
perturbations began to convert into curvature ones, and this
conversion will complete in two possible cases, namely, either when
the curvaton dominates the universe, or decay as well, whichever is
earlier \footnote{Actually, besides the standard model radiation,
the curvaton can also decay into other products such as dark
radiation or dark matter. In that case, there may have residual
isocurvature perturbations and the non-Gaussianities may also be
different \cite{Mazumdar:2011xe} (see also \cite{Mazumdar:2010sa}
for a review). In this paper we assume that curvaton decays to
standard model radiation for simplicity. We thank A. Mazumdar for
point out this for us via private communication.}. We assume that
the conversion as well as curvaton decay happens instantaneously, so
that we can separately consider the curvature perturbations for each
component (radiation and curvaton) and just use Eq.
(\ref{curvature}) to calculate the final total curvature
perturbation of the universe. From Eqs. (\ref{curvatoncurv}) and
(\ref{curvature}) we know that the final curvature perturbation can
be expressed as:
\begin{eqnarray}\label{zeta2}
\zeta=\frac{\delta\rho_\sigma}{4\rho_r+3(\rho_\sigma+p_\sigma)}~,
\end{eqnarray} where the density perturbation $\delta\rho_\sigma$
can furtherly be expanded with respect to $\delta\sigma$. The linear
order of $\delta\rho_\sigma$ is given by:
\begin{eqnarray}
\delta^{(1)}\rho_\sigma &\simeq& \rho_{\sigma,\sigma}\delta\sigma\nonumber\\
&=&
\label{deltarho1}(2K_{,X\sigma}X-K_{,\sigma}+3HG_{,X\sigma}\dot\sigma_0^3-2G_{,\sigma\sigma}
X)\delta\sigma~,
\end{eqnarray}
while the second order of $\delta\rho_\sigma$ reads:
\begin{eqnarray}
\delta^{(2)}\rho_\sigma &\simeq&
\frac{1}{2}\rho_{\sigma,\sigma\sigma}\delta\sigma^2 \nonumber\\
\label{deltarho2}&=& \frac{1}{2}
(2K_{,X\sigma\sigma}X-K_{,\sigma\sigma}+3HG_{,X\sigma\sigma}\dot\sigma_0^3
-2G_{,\sigma\sigma\sigma} X)\delta\sigma^2~
\end{eqnarray} respectively.

Now we can consider the two cases separately. First, if the curvaton
dominates the energy density before decays, we can use Eqs.
(\ref{curvatoncurv}), (\ref{curvature1}) as well as
(\ref{deltarho1}) to have the final curvature perturbation as:
\begin{eqnarray}
\zeta^{(I)} &\simeq&
\frac{\delta^{(1)}\rho_\sigma}{3(\rho_\sigma+p_\sigma)}
\simeq\frac{\rho_{\sigma,\sigma}}{3(\rho_\sigma+p_\sigma)}\delta\sigma
\nonumber\\
&=&
\frac{1}{3}\bigg(\frac{K_{,X\sigma}\dot{\sigma}_0^{2}-K_{,\sigma}+3HG_{,X\sigma}\dot{\sigma}_0^{3}-G_{,\sigma\sigma}\dot{\sigma}_0^{2}}
{K_{,X}\dot{\sigma}_0^{2}+3HG_{,X}\dot{\sigma}_0^{3}-2G_{,\sigma}\dot{\sigma}_0^{2}-G_{,X}\ddot{\sigma}_0\dot{\sigma}_0^{2}}\bigg)\delta\sigma~,
\end{eqnarray} and the power spectrum of curvature perturbation is: \begin{eqnarray}
{\cal P}^{(I)}_\zeta &\equiv& \frac{k^3}{2\pi^2}|\zeta|^2 \nonumber\\
\label{curvspectrum1}&=&\frac{1}{9}\bigg(\frac{H^2_\ast}{4\pi^2
c_s^3\mathcal{D}}\bigg)\bigg(\frac{K_{,X\sigma}\dot{\sigma}_0^{2}-K_{,\sigma}+3HG_{,X\sigma}\dot{\sigma}_0^{3}-G_{,\sigma\sigma}\dot{\sigma}_0^{2}}
{K_{,X}\dot{\sigma}_0^{2}+3HG_{,X}\dot{\sigma}_0^{3}-2G_{,\sigma}\dot{\sigma}_0^{2}-G_{,X}\ddot{\sigma}_0\dot{\sigma}_0^{2}}\bigg)^2~,
\end{eqnarray} where we also made use of our previous results
(\ref{spectrum2}). Second, if the curvaton decays before its
dominance, it will only contribute part of the energy density of the
universe with some fraction $r$ as defined before. Using Eq.
(\ref{curvature2}), in this case the curvature perturbation is given
by
\begin{eqnarray}
\zeta^{(II)}&\simeq&\frac{r}{4}\frac{\delta^{(1)}\rho_\sigma}{\rho_\sigma}
\simeq\frac{r}{4}\frac{\rho_{\sigma,\sigma}}{\rho_\sigma}\delta\sigma
\nonumber\\
&=&
\frac{r}{4}\bigg(\frac{K_{,X\sigma}\dot{\sigma}_0^{2}-K_{,\sigma}+3HG_{,X\sigma}\dot{\sigma}_0^{3}-G_{,\sigma\sigma}\dot{\sigma}_0^{2}}
{K_{,X}\dot{\sigma}_0^{2}-K+3HG_{,X}\dot{\sigma}_0^{3}-G_{,\sigma}\dot{\sigma}_0^{2}}\bigg)\delta\sigma~,
\end{eqnarray} and the curvature perturbation power spectrum is:
\begin{eqnarray}
{\cal P}^{(II)}_\zeta
&=&\frac{r^2}{16}\bigg(\frac{K_{,X\sigma}\dot{\sigma}_0^{2}-K_{,\sigma}+3HG_{,X\sigma}\dot{\sigma}_0^{3}-G_{,\sigma\sigma}\dot{\sigma}_0^{2}}
{K_{,X}\dot{\sigma}_0^{2}-K+3HG_{,X}\dot{\sigma}_0^{3}-G_{,\sigma}\dot{\sigma}_0^{2}}\bigg)^2\frac{k^3}{2\pi^2}|\delta\sigma|^2\nonumber\\
\label{curvspectrum2}&=&\frac{r^2}{16}\bigg(\frac{H^2_\ast}{4\pi^2
c_s^3\mathcal{D}}\bigg)\bigg(\frac{K_{,X\sigma}\dot{\sigma}_0^{2}-K_{,\sigma}+3HG_{,X\sigma}\dot{\sigma}_0^{3}-G_{,\sigma\sigma}\dot{\sigma}_0^{2}}
{K_{,X}\dot{\sigma}_0^{2}-K+3HG_{,X}\dot{\sigma}_0^{3}-G_{,\sigma}\dot{\sigma}_0^{2}}\bigg)^2~.
\end{eqnarray} Here the supscripts $(I)$ and $(II)$ refer to the two
cases respectively.

From Eqs. (\ref{curvspectrum1}) and (\ref{curvspectrum2}) one can
see that comparing to usual curvaton case, the amplitude of
curvature perturbation will also be modified with a different
prefactor, which contains contribution from the high derivative term
$G(\sigma,X)$. So one can naturally expect G-Curvaton model has some
feature that could be distinguished from usual K-essence curvaton
models.

\subsection{Non-Gaussianities generated by G-Curvaton}

In this paragraph, we extend our study to non-linear perturbations
of our model, i.e., non-Gaussianities. First of all, we consider the
local type non-Gaussianities generated by the curvaton model. For
local type, the nonlinear parameter $f^{local}_{NL}$ has been
defined in (\ref{zeta}). From Eqs. (\ref{zeta2}-\ref{deltarho2}) and
using the $\delta N$ formalism (\ref{deltan}), we can also express
$f^{local}_{NL}$ in terms of energy density as:
\begin{eqnarray}
f^{local}_{NL}\Big|_\zeta=\frac{5}{6}[4\rho_r+3(\rho_\sigma+p_\sigma)]\frac{\rho_{\sigma,\sigma\sigma}}{\rho_{\sigma,\sigma}^2}~.
\end{eqnarray}

Now we consider the two cases separately. For the first case where
curvaton dominates the energy density before decays, one gets:
\begin{eqnarray}
f^{local}_{NL}\Big|_\zeta^{(I)}&\simeq&\frac{5}{2}\frac{(\rho_\sigma+p_\sigma)\rho_{\sigma,\sigma\sigma}}{\rho_{\sigma,\sigma}^2}\nonumber\\
\label{local1}&=&\frac{5}{2}\bigg[\frac{(K_{,X}\dot{\sigma}_0^{2}+3HG_{,X}\dot{\sigma}_0^{3}-2G_{,\sigma}\dot{\sigma}_0^{2}-G_{,X}\ddot{\sigma}_0\dot{\sigma}_0^{2})
(K_{,X\sigma\sigma}\dot{\sigma}_0^{2}-K_{,\sigma\sigma}+3HG_{,X\sigma\sigma}\dot{\sigma}_0^{3}
-G_{,\sigma\sigma\sigma}\dot{\sigma}_0^{2})}
{(K_{,X\sigma}\dot{\sigma}_0^{2}-K_{,\sigma}+3HG_{,X\sigma}\dot{\sigma}_0^{3}-G_{,\sigma\sigma}\dot{\sigma}_0^{2})^2}\bigg]~,
\end{eqnarray}
and for the second case where the curvaton decays and never
dominates the energy density, we have
\begin{eqnarray}
f^{local}_{NL}\Big|_\zeta^{(II)}&\simeq&\frac{10}{3r}\frac{\rho_\sigma\rho_{\sigma,\sigma\sigma}}{\rho_{\sigma,\sigma}^2}
\nonumber\\
\label{local2}&=&
\frac{10}{3r}\bigg[\frac{(K_{,X}\dot{\sigma}_0^{2}-K+3HG_{,X}\dot{\sigma}_0^{3}-G_{,\sigma}\dot{\sigma}_0^{2})
(K_{,X\sigma\sigma}\dot{\sigma}_0^{2}-K_{,\sigma\sigma}+3HG_{,X\sigma\sigma}\dot{\sigma}_0^{3}
-G_{,\sigma\sigma\sigma}\dot{\sigma}_0^{2})}
{(K_{,X\sigma}\dot{\sigma}_0^{2}-K_{,\sigma}+3HG_{,X\sigma}\dot{\sigma}_0^{3}-G_{,\sigma\sigma}\dot{\sigma}_0^{2})^2}\bigg]~,
\end{eqnarray}
respectively.

Then we turn to the nonlocal type of non-Gaussianities. The nonlocal
type of non-Gaussianities is more complicated, as described in Eqs.
(\ref{nonlocal1}-\ref{nonlocal3}). The interaction hamiltonian is
${\cal H}^p_{int}=-{\cal L}_3$, where ${\cal L}_3$ is the 3-rd order
perturbed Lagrangian. Starting from action (\ref{actiongc}) and
after straightforward but rather tedious calculation, we get the
interaction hamiltonian at the leading order with respect to slow
roll parameters:
\begin{eqnarray}\label{hamiltonian}
H_{int}^{p}&\supset&\int\frac{d^{3}k_{1}d^{3}k_{2}d^{3}k_{3}}{(2\pi)^{9}}(2\pi)^{3}\delta^{3}(\mathbf{k}_{1}+\mathbf{k}_{2}+\mathbf{k}_{3})
\{+aL_{1}(\mathbf{k}_{2}\cdot\mathbf{k}_{3})\dot{\delta\sigma}(t,\mathbf{k}_{1})\dot{\delta\sigma}(t,\mathbf{k}_{2}){\delta\sigma}(t,\mathbf{k}_{3})
+a^{3}L_{2}\dot{\delta\sigma}(t,\mathbf{k}_{1})\dot{\delta\sigma}(t,\mathbf{k}_{2})\dot{\delta\sigma}(t,\mathbf{k}_{3})\nonumber\\
&&+aL_{3}(\mathbf{k}_{2}\cdot\mathbf{k}_{3})\dot{\delta\sigma}(t,\mathbf{k}_{1}){\delta\sigma}(t,\mathbf{k}_{2}){\delta\sigma}(t,\mathbf{k}_{3})
+a^{-1}L_{4}(\mathbf{k}_{1}\cdot\mathbf{k}_{2})\mathbf{k}_{3}^{2}{\delta\sigma}(t,\mathbf{k}_{1}){\delta\sigma}(t,\mathbf{k}_{2}){\delta\sigma}(t,\mathbf{k}_{3})\}~,
\end{eqnarray} where we have defined \begin{eqnarray}
L_{1}&=&G_{,XX0}\dot{\sigma}_{0}^{2}~,\\
L_{2}&=&\frac{1}{2}(5HG_{,XX0}\dot{\sigma}_{0}^{2}+2HG_{,X0}+\frac{1}{3}K_{,XXX0}\dot{\sigma}_{0}^{3}+HG_{,XXX0}\dot{\sigma}_{0}^{4}+K_{,XX0}\dot{\sigma}_{0})~,\\
L_{3}&=&-\frac{1}{2}(K_{,XX0}\dot{\sigma}_{0}+2G_{,XX0}\dot{\sigma}_{0}\ddot{\sigma}_{0}+3HG_{,XX0}\dot{\sigma}_{0}^{2}+4HG_{,X0})~,\\
L_{4}&=&-\frac{1}{2}G_{,X0}~.
\end{eqnarray}

Moreover, from the solution of the canonical variable $u_k$ we can
obtain the mode solution of the perturbation variable (in conformal
time) $\delta\sigma(\tau,k)$ as:
\begin{eqnarray}\label{mode} \delta\sigma(\tau,{\bf k})=q(\tau,{\bf
k})a_{\bf k}+q^\ast(\tau,{\bf k})a_{-{\bf
k}}^\dagger~,~~~q(\tau,{\bf
k})=\frac{iH}{\sqrt{2\mathcal{D}c_{s}^{3}k^{3}}}(1+ic_{s}k\tau)e^{-ic_{s}k\tau}~,\end{eqnarray}
where $a_{\bf k}$ and $a_{-{\bf k}}^\dagger$ are production and
annihilation operators respectively. By substituting Eq.
(\ref{hamiltonian}-\ref{mode}) into (\ref{nonlocal2}) , one can get
the result of 3-point correlation function
$\langle|\delta\sigma(k_1)\delta\sigma(k_2)\delta\sigma(k_3)|\rangle$.
Note that when carrying out integrations with respect to conformal
time $\tau$, we assume that the $L_i$'s $(i=1,2,3,4)$ are all slow
varying and can be roughly taken out of the integrations, which
greatly simplifies our calculation. Similar assumptions has been
made when calculating more general single field inflation with
nonminimal coupling to Gravity as well as Gauss-Bonnet terms
\cite{DeFelice:2011zh} though more rigid calculation with nonminimal
coupling only was performed in \cite{Qiu:2010dk}. The final shape of
$\langle|\delta\sigma(k_1)\delta\sigma(k_2)\delta\sigma(k_3)|\rangle$
than reads:
\begin{eqnarray}
\mathcal{B}_{\delta\sigma}(k_{1},k_{2},k_{3})&=&\frac{3L_{1}H^{6}}
{\mathcal{D}^{3}c_{s}^{8}k_{1}k_{2}k_{3}K^{3}}-\frac{3L_{2}H^{5}}{\mathcal{D}^{3}c_{s}^{6}k_{1}k_{2}k_{3}K^{3}}+
\frac{L_{3}H^{5}}{4\mathcal{D}^{3}c_{s}^{8}k_{1}^{3}k_{2}^{3}k_{3}^{3}K^{3}}(6k_{1}k_{2}k_{3}\sum_{i}k_{i}^{3}+2\sum_{i\neq
j}k_{i}^{2}k_{j}^{4}+3\sum_{i\neq
j}k_{i}k_{j}^{5}\nonumber\\
&&+\sum_{i}k_{i}^{6})+\frac{L_{4}H^{6}}{2\mathcal{D}^{3}c_{s}^{10}k_{1}^{3}k_{2}^{3}k_{3}^{3}K^{4}}(2\sum_{i\neq
j}k_{i}^{2}k_{j}^{5}-7\sum_{i\neq j}k_{i}^{4}k_{j}^{3}+4\sum_{i\neq
j}k_{i}k_{j}^{6}-18k_{1}^{2}k_{2}^{2}k_{3}^{2}\sum_{i}k_{i}\nonumber\\
&&-4k_{1}k_{2}k_{3}\sum_{i\neq
j}k_{i}^{3}k_{j}-24k_{1}k_{2}k_{3}\sum_{i>j}k_{i}^{2}k_{j}^{2}+12k_{1}k_{2}k_{3}\sum_{i}k_{i}^{4}+\sum_{i}k_{i}^{7})~,\end{eqnarray}
and the corresponding non-linear estimator $f_{NL}$ can be
calculated according to Eq. (\ref{nonlocal3}). As an example, here
we consider the equilateral case where $k_1=k_2=k_3$, in which the
estimator then reduces to:
\begin{eqnarray}
f_{NL}^{equil}\Big|_{\delta\sigma}&=&\frac{10}{27}(\frac{L_{1}H^{2}}{9\mathcal{D}c_{s}^{2}}-\frac{L_{2}H}{9\mathcal{D}}+\frac{17L_{3}H}{36\mathcal{D}c_{s}^{2}}-\frac{13L_{4}H^{2}}{18\mathcal{D}c_{s}^{4}})~,
\end{eqnarray}
while using $\delta N$ formalism (\ref{deltan}), the nonlinear
estimator for curvature perturbation reads: \begin{eqnarray}
f_{NL}^{equil}\Big|_{\zeta}=N_{,\sigma}f_{NL}^{equil}\Big|_{\delta\sigma}~.
\end{eqnarray}

From Eqs. (\ref{zeta2}) and (\ref{deltarho1}), we can easily get
$N_{,\sigma}=\rho_{\sigma,\sigma}/[4\rho_r+3(\rho_\sigma+p_\sigma)]$,
which gives \begin{eqnarray}
f_{NL}^{equil}\Bigl|_{\zeta}^{(I)}&=&\frac{\rho_{\sigma,\sigma}}{3(\rho_{\sigma}+p_{\sigma})}f_{NL}^{equil}\Bigl|_{\delta\sigma}\nonumber\\
\label{equil1}&=&\frac{10\rho_{\sigma,\sigma}}{81(\rho_{\sigma}+p_{\sigma})}(\frac{L_{1}H^{2}}{9\mathcal{D}c_{s}^{2}}-\frac{L_{2}H}{9\mathcal{D}}+\frac{17L_{3}H}{36\mathcal{D}c_{s}^{2}}-\frac{13L_{4}H^{2}}{18\mathcal{D}c_{s}^{4}})
\end{eqnarray} for the case of which curvaton dominates before decays,
and \begin{eqnarray}
f_{NL}^{equil}\Bigl|_{\zeta}^{(II)}&=&\frac{r\rho_{\sigma,\sigma}}{4\rho_{\sigma}}f_{NL}^{equil}\Bigl|_{\delta\sigma}\nonumber\\
\label{equil2}&=&\frac{5r\rho_{\sigma,\sigma}}{54\rho_{\sigma}}(\frac{L_{1}H^{2}}{9\mathcal{D}c_{s}^{2}}-\frac{L_{2}H}{9\mathcal{D}}+\frac{17L_{3}H}{36\mathcal{D}c_{s}^{2}}-\frac{13L_{4}H^{2}}{18\mathcal{D}c_{s}^{4}})
\end{eqnarray} for the case of which curvaton decays and never dominates.

\subsection{A concrete G-Curvaton model}
In previous parts of this section, we have presented the whole
process of how our G-Curvaton model works, including its background
evolution, scale-invariant isocurvature perturbation generation,
curvature perturbation conversion as well as different types of
non-Gaussianities. However, due to the involvement of our model,
especially containing high derivative terms, the above general
analysis can only be rather qualitative. Moreover, unlike the usual
curvaton model, there are many uncertainties in our model with
general form (\ref{actiongc}) and it can have various decaying
mechanisms, each of which may have different results for curvature
perturbations and non-Gaussianities. In order to make things more
specific, it is necessary to focus on some explicit models to see
how our models can be different from the usual curvaton models.

Before constructing models, let's investigate how many constraints
we have to consider for our model. First of all, as was mentioned in
previous section, a curvaton model must have light mass, that is,
${\cal M}_{eff}^{2}\ll H^2$. In slow roll approximation, we have
$\dot\sigma\ll M_{pl}^2$, so the first term of Eq. (\ref{mass}) can
then be neglected, which makes ${\cal M}_{eff}^{2}\simeq
-K_{,\sigma\sigma}+3HG_{,\sigma\sigma}\dot\sigma$. Therefore, if we
choose $|K_{,\sigma\sigma}|$ and $|G_{,\sigma\sigma}|$ to be small
enough, it will be safe for curvaton. Another way is both
$K_{,\sigma\sigma}$ and $3H\dot\sigma G_{,\sigma\sigma}$ may be
large in amplitude, but are of similar value and opposite sign. In
this case, they can be canceled to have a relatively small value,
which may need some fine tuning in the model. The second constraint
comes from the observations. The current observation data gives very
tight constraint on the amplitude of the curvature perturbations,
for example, the WMAP-7 measurement of the CMB quadrupole anisotropy
requires ${\cal P}_\zeta\sim2.4\times10^{-9}$ \cite{Komatsu:2010fb}.
In order for the amplitude of the curvature perturbation in
(\ref{curvspectrum1}) and (\ref{curvspectrum2}) to meet the data,
one can furtherly constrain the form of Lagrangian of our model,
namely $K(\sigma,X)$ and $G(\sigma,X)$ and their field dependence.

Taking account to both constraints from above, we can consider that
our model may have the Lagrangian with form of:
\begin{eqnarray}\label{model}
K(\sigma,X)=X-V(\sigma)~,~~~G(\sigma,X)=-g(\sigma)X
\end{eqnarray}
as an example. Here we require both $V_{,\sigma\sigma}$ and
$g_{,\sigma\sigma}$ be small enough to give rise to small effective
mass needed for curvaton. For background evolution, from eqs.
(\ref{rho}), (\ref{p}) and (\ref{bgeom}), we have the following
equations:
\begin{eqnarray}
&\rho_\sigma=X(1-6gH\dot{\sigma}_0+g_{,\sigma}\dot{\sigma}_0^{2})+V~,&\\
&\rho_\sigma+p_\sigma=2X(1+g\ddot{\sigma}_0-3gH\dot{\sigma}_0+g_{,\sigma}\dot{\sigma}_0^{2})~,&\\
&\ddot{\sigma}_0+3H\dot{\sigma}_0+2g_{,\sigma}\dot{\sigma}_0^{2}\ddot{\sigma}_0+\frac{1}{2}g_{,\sigma\sigma}\dot{\sigma}_0^{4}
-3g\dot{H}\dot{\sigma}_0^{2}-6gH\dot{\sigma}_0\ddot{\sigma}_0-9gH^{2}\dot{\sigma}_0^{2}+V_{,\sigma}=0~.&
\end{eqnarray}
For perturbations, from Eqs. (\ref{c}) and (\ref{d}) we can get:
\begin{eqnarray} {\cal C}=1-2g\ddot\sigma_0-4gH\dot\sigma_0~,~~~
{\cal D}=1+4g_{,\sigma}X-6gH\dot\sigma_0~
\end{eqnarray} for our model, which can give the sound speed squared $c_s^2={\cal C}/{\cal
D}$. One can also get the power spectrum and non-Gaussianities of
curvature perturbations by making use of the explicit form
(\ref{model}) in the corresponding formulae that has been derived in
the above sections.

For later convenience, let us first introduce some more
``slow-variation" parameters, namely
\begin{eqnarray}\label{slowroll2}
\eta\equiv\frac{\ddot\sigma_0}{H\dot\sigma_0}~,~~~\alpha\equiv\frac{\dot
g}{gH}~,~~~\beta&\equiv&\frac{\dot{g}_{,\sigma}}{g_{,\sigma}H}~,~~~\gamma\equiv\frac{\dot{g}_{,\sigma\sigma}}{g_{,\sigma\sigma}H}~.
\end{eqnarray}
We can always choose the form of $g(\sigma)$ such that the last
three parameters $|\alpha|,~|\beta|,~|\gamma| \ll 1$ all the time.
However, $|\eta|$ can only be small when the field remains
slow-rolling. This is important because in curvaton scenario,
curvature perturbations are produced {\it after} the end of
inflation, and in that case, the slow-rolling of the curvaton field
is not always satisfied.

Now let's turn on to the two cases of generating curvature
perturbations one by one. In the first case the curvature
perturbation is generated when curvaton dominates the universe. That
requires the decaying of the curvaton is slower than that of
radiation which is transferred from inflaton. For instance, it is
reasonable to assume that the curvaton field is still slow-rolling,
and exit in a few number of efolds, so it will not lead to another
period of rapid acceleration \cite{Kobayashi:2010cm}. Substituting
(\ref{model}) into Eqs. (\ref{curvspectrum1}), (\ref{local1}) and
(\ref{equil1}) and letting all the slow-variation parameters defined
in (\ref{slowroll2}) be small, we can get:
\begin{eqnarray}
{\cal
P}_{\zeta}^{(I)}&\simeq&(\frac{H_{\ast}^{2}}{4\pi^{2}c_{s}^{3}\mathcal{D}})(H/\dot\sigma_0)^{2}~,\\
f_{NL}^{local}\Bigl|_{\zeta}^{(I)}&\simeq&\frac{5}{6}[\frac{(3\alpha
gH\dot{\sigma}_0-6\epsilon
gH\dot{\sigma}_0+\epsilon)}{(1-3gH\dot{\sigma}_0)}]~,\\
f_{NL}^{equil}\Bigl|_{\zeta}^{(I)}&\simeq&\frac{10gH\dot{\sigma}_0}{243\mathcal{D}}(\frac{(3-\epsilon)gH\dot{\sigma}_0-1}{1-3gH\dot{\sigma}_0})(1+\frac{17}{2c_{s}^{2}}-\frac{13}{4c_{s}^{4}})(H/\dot\sigma_0)^{2}~,
\end{eqnarray}

Similarly, in the second case the curvature perturbation is
generated when the curvaton decays and becomes equilibrium with
radiation, and thus the energy density of curvaton has the same
scaling as that of radiation with the ratio $r=\rho_\sigma/\rho_r$.
Again using (\ref{model}) with Eqs. (\ref{curvspectrum2}),
(\ref{local2}) and (\ref{equil2}), letting $\alpha$, $\beta$ and
$\gamma$ be small but retaining $\eta$ for the reason given above,
we can get: \begin{eqnarray} {\cal
P}_{\zeta}^{(II)}&\simeq&\frac{(1+r)^{2}}{16}(\frac{H_{\ast}^{2}}{4\pi^{2}c_{s}^{3}\mathcal{D}})[(3-\epsilon+2\eta)gH\dot{\sigma}_0-\frac{1}{3}(\eta+3)]^{2}(H/\dot\sigma_0)^{-2}~,\\
f_{NL}^{local}\Bigl|_{\zeta}^{(II)}&\simeq&\frac{20\epsilon}{(r+1)}[\frac{(-6(3-\epsilon+2\eta)gH\dot{\sigma}_0+(\eta+3))}{(3(3-\epsilon+2\eta)gH\dot{\sigma}_0-(\eta+3))^{2}}](H/\dot\sigma_0)^{2}~,\\
f_{NL}^{equil}\Bigl|_{\zeta}^{(II)}&\simeq&\frac{5(r+1)gH\dot\sigma_0}{486\mathcal{D}}[(3-\epsilon+2\eta)gH\dot{\sigma}_0-\frac{1}{3}(\eta+3)](1+\frac{17}{2c_{s}^{2}}-\frac{13}{4c_{s}^{4}})~.
\end{eqnarray}

From above we can see that another variable that appears repeatedly
is $gH\dot\sigma_0$. Defining $x:= gH\dot\sigma_0$, and in order to
avoid ghost and gradient instability, $x$ should be smaller than
unity. We consider two asymoptotic cases: i) $x$ could be positive
or negative, with $|x|\ll 1$ and ii) $x$ is negative, with $|x|\gg
1$, which means that the Galileon term plays a unimportant/important
role at the end of inflation respectively. The spectrum and
non-Gaussianities of the
curvature perturbation can be reduced as:\\
i) $|x|\ll 1$: $c_s^2\simeq 1$
\begin{eqnarray}
&&{\cal
P}_{\zeta}^{(I)}\simeq(\frac{H_{\ast}^{2}}{4\pi^{2}})(H/\dot\sigma_0)^{2}~,~f_{NL}^{local}\Bigl|_{\zeta}^{(I)}\simeq\frac{5}{6}\epsilon~,~f_{NL}^{equil}\Bigl|_{\zeta}^{(I)}\simeq-\frac{125x}{486}(H/\dot\sigma_0)^{2}~,\\&&{\cal
P}_{\zeta}^{(II)}\simeq\frac{(1+r)^{2}}{144}(\frac{H_{\ast}^{2}}{4\pi^{2}})\frac{(\eta+3)^{2}}{(H/\dot\sigma_0)^{2}}~,~f_{NL}^{local}\Bigl|_{\zeta}^{(II)}\simeq\frac{20\epsilon(H/\dot\sigma_0)^{2}}{(r+1)(\eta+3)}~,~f_{NL}^{equil}\Bigl|_{\zeta}^{(II)}\simeq-\frac{125(r+1)x}{5832}(\eta+3)~,
\end{eqnarray} ii) $|x|\gg 1$: $c_s^2\simeq \frac{\eta+2}{3}$ ($\eta$ is small for Case I)
\begin{eqnarray}
&&{\cal
P}_{\zeta}^{(I)}\simeq\frac{1}{4|x|}\sqrt{\frac{3}{2}}(\frac{H_{\ast}^{2}}{4\pi^{2}})(H/\dot\sigma_0)^{2}~,~f_{NL}^{local}\Bigl|_{\zeta}^{(I)}\simeq\frac{5}{6}(2\epsilon-\alpha)~,~f_{NL}^{equil}\Bigl|_{\zeta}^{(I)}\simeq-\frac{35(3-\epsilon)}{17496}(H/\dot\sigma_0)^{2}~,\\
&&{\cal
P}_{\zeta}^{(II)}\simeq\frac{(1+r)^{2}|x|}{32}\sqrt{\frac{3}{(\eta+2)^{3}}}(\frac{H_{\ast}^{2}}{4\pi^{2}})\frac{(3-\epsilon+2\eta)^{2}}{(H/\dot\sigma_0)^{2}}~,~f_{NL}^{local}\Bigl|_{\zeta}^{(II)}\simeq\frac{40\epsilon(H/\dot\sigma_0)^{2}}{3(r+1)(3-\epsilon+2\eta)|x|}~,\nonumber\\
&&f_{NL}^{equil}\Bigl|_{\zeta}^{(II)}\simeq\frac{5(r+1)(3-\epsilon+2\eta)(103+118\eta+4\eta^{2})}{11664|x|(2+\eta)^{2}}~,\end{eqnarray}
respectively.

From the above results we can have a couple of comments on the
perturbations of our G-Curvaton model. Besides the slow variation
parameters which are roughly of order 1, now we have three more free
parameters, namely the value of $x$, $H$ and $\dot\sigma_0$ at the
end of inflation, to determine ${\cal P}_\zeta$ and $f_{NL}$.
Considering the constraints from CMB that the amplitude of power
spectrum to be nearly $10^{-9}$, we still have large parameter space
to have considerable large non-Gaussianities. For instance, for Case
I the curvaton field is still slow-rolling at the end of inflation,
where $|\dot\sigma_0|$ is small compared to $H$. If it satisfies
$|\dot\sigma_0|\sim10^{-3}H$, then from the power spectrum we have
$H\sim10^{-8}$, which is slightly lower than that of chaotic
inflation. For subcase where $|x|\ll 1$, we can have
$f_{NL}^{equil}\sim10^2$ just by requiring $|g|\sim10^{-13}$. For
subcase where $|x|\gg 1$, however, we can have
$f_{NL}^{equil}\sim10^2$ by requiring $|g|\sim10^{18}$. Moreover, in
both cases $f_{NL}^{local}$ remains of order unity. For Case II, the
energy density of $\sigma$ is comparable to $H$, which roughly gives
$\dot\sigma_0^2[1+{\cal O}(1)x]\sim H^2$. In subcase $|x|\ll 1$, we
have $|\dot\sigma_0|\sim H$, which gives $H\sim 10^{-5}$ in order to
meet the constraint on power spectrum. This in turn gives small
$f_{NL}^{equil}$ with $f_{NL}^{local}$ roughly of ${\cal O}(10)$. In
subcase $|x|\gg 1$, however, we obtain $|g\dot\sigma_0^3|\sim H$. If
furtherly it satisfies, i.e., $|\dot\sigma_0|\sim10^{-1}H$, it will
lead to $H\sim10^{-2}$ from constraint on power spectrum, which in
turn gives $\dot\sigma_0\sim10^{-3}$ as well as $|g|\sim10^7$. Then
the nonlinear estimator $f_{NL}^{local}$ is roughly of ${\cal
O}(10)$, and $f_{NL}^{equil}$ becomes of order unity\footnote{In our
analysis we also made the approximation of $H_\ast\sim H$, where
$H_\ast$ is the value of Hubble parameter when the fluctuations of
$\sigma$ exits the horizon during inflation.}. Moreover, due to the
non-linear effects of the Galileon term $G(X,\sigma)\Box\sigma$ in
our model, it can be expected that large tensor-to-scalar-ratio can
be generated as well as large non-Gaussianities
\cite{Kobayashi:2010cm}, which can be distinguishable from the
standard curvaton model, of which large non-Gaussianities are also
accompanied with lower energy scale of inflation, which leads to
small tensor-to-scalar ratio (cf. the second paper in
\cite{Sasaki:2006kq}).

\section{Conclusion and Discussions}
In this paper, we investigated G-Curvaton scenario, where the
curvaton field is acted by Galileon action. This opens a new access
of curvaton scenarios that the fluctuations could be affected by
nonlinear terms. After reviewing the standard curvaton mechanism, we
started from the action of Galileon field and calculated the
spectrum of the field perturbation. The power spectrum is suppressed
by $\mathcal{D}$ given in (\ref{d}), rather than $K_{,X}$ in the
normal single field case. We studied the generation of curvature
perturbations in both two possible cases, and apart from curvature
perturbation, we obtained non-Gaussianities of both local and
non-local types.

We have shown, in this work, that there is large possibilities for
G-Curvaton to have consistent power spectrum and large
non-Gaussianities. We presented a concrete model of G-Curvaton as an
illustration. With proper choice of the parameters, the nonlinear
estimator could be made of ${\cal O}(10^2)$. However, because of the
large parameter space, the conclusion is still highly model
dependent. We can expect future observational data to have more
rigid constraints on the G-Curvaton scenario, for example, if future
observations can observe large non-local non-Gaussianity compared to
local one, more or less we can say that it {\it might} due to some
nonlinear effects such as Galileon.

Moreover, since the Galileon field can violate the NEC without the
ghost and gradient instabilities, G-Curvaton can naturally
incorporate a model of curvaton with NEC violation, which might be
interesting for studying. Here, we focus on the inflationary
background, however, in principle, G-Curvaton can also be embedded
into alternative models to inflation, which will make the curvature
perturbation induced in corresponding model have more fruitful
predictions. For phenomenological aspect, there are also a couple of
implications that merits thinking of: one example is that large
non-Gaussianities could be accompanied with large tensor-to-scalar
ratio as well, which is different from usual curvaton models and
thus can be used as a distinguishment observationally; another is
that due to the non-linear effects from Galileon term, the decaying
process of the curvaton might also be modified, which may in turn
change the speed and amount of the generated products during
reheating \cite{LevasseurPerreault:2011mw}\footnote{We thank the
referee for reminding us these two points.}. We hope that some
upcoming works could gain more clear insights into G-Curvaton
scenario.

\vspace{16mm}

\begin{acknowledgements}
T.Q. thanks Prof. Xinmin Zhang for the hospitality during his visit
in Institute of High Energy Physics, China, where most of this work
has been done, and also in College of Physical Sciences, Graduate
University of Chinese Academy of Sciences. H.W. thanks Jun Zhang and
Zhi-Guo Liu for helpful discussions. Y.S.P. is supported in part by
NSFC under Grant No:10775180, 11075205, in part by the Scientific
Research Fund of GUCAS(NO:055101BM03), in part by National Basic
Research Program of China, No:2010CB832804.
\end{acknowledgements}

\end{document}